\documentclass[a4paper,10pt]{article}
\usepackage{stmaryrd}
\usepackage{amsfonts}
\usepackage{bbm}
\usepackage{amscd}
\usepackage{mathrsfs}
\usepackage{latexsym,amssymb,amsmath,amscd,amscd,amsthm,amsxtra,xypic}
\usepackage[dvips]{graphicx}

\newtheorem{thm}{Theorem}[section]
\newtheorem{lem}[thm]{Lemma}
\newtheorem{cor}[thm]{Corollary}
\newtheorem{pro}[thm]{Proposition}
\newtheorem{ex}[thm]{Example}
\newtheorem{rmk}[thm]{Remark}
\newtheorem{defi}[thm]{Definition}

\newcommand{\gl }{{\mathfrak{gl} } }

\newcommand{\lon }{\,\rightarrow\,}
\newcommand{\be }{\begin{eqnarray*}}
\newcommand{\ee }{\end{eqnarray*}}

\newcommand{\defbe}{\triangleq}
\newcommand{\pf}{\noindent{\bf Proof.}\ }

\newcommand{\Real}{\mathbb R}

\newcommand{\huaE}{\mathcal{E}}

\newcommand{\CWM}{C^{\infty}(M)}

\newcommand{\XM}{\mathcal{X}(M)}

\newcommand{\set}[1]{\left\{#1\right\}}

\newcommand{\pairing}[1]{\left\langle #1\right\rangle}

\newcommand{\pibracket}[1]{\left [ #1\right ]_{\pi}}

\newcommand{\frkd}{\mathfrak d}

\newcommand{\frkr}{\mathfrak r}

\newcommand{\frky}{\mathfrak y}
\newcommand{\frkz}{\mathfrak z}

\newcommand{\frkL}{\mathfrak L}

\newcommand{\ppairingE}[1]{\left ( #1\right )_E}

\def\gpd{\,\lower1pt\hbox{$\longrightarrow$}\hskip-.24in\raise2pt
         \hbox{$\longrightarrow$}\,}

\def\qed{\hfill ~\vrule height6pt width6pt depth0pt}

\newcommand{\LieDerivation}{\frkL}

\newcommand{\pisharp}{\pi}

\newcommand{\half}{\frac{1}{2}}

\newcommand{\conpairing}[1]{\left\langle  #1\right\rangle }
\newcommand{\Courant}[1]{\left\llbracket  #1\right\rrbracket }
\newcommand{\Dorfman}[1]{\left \{ #1\right \} }

\newcommand{\jet}{\mathfrak{J}}
\newcommand{\jetd}{\mathbbm{d}}
\newcommand{\dev}{\mathfrak{D}}
\newcommand{\TE}{TE}

\newcommand{\EStar}{{E^*}}

\newcommand{\pE}{p_E}

\newcommand{\TStarM}{T^*M}

\newcommand{\pie}{^\prime}
\newcommand{\Id}{\mathbf{1}}

\newcommand{\e}{\mathbbm{e}}
\newcommand{\p}{\mathbbm{p}}
\newcommand{\id}{\mathbbm{i}}
\newcommand{\jd}{\alpha}
\newcommand{\jE}{\mathfrak{J}{E}}
\newcommand{\dE}{\mathscr{D}{E}}

\newcommand{\dM}{\mathrm{d}}
\newcommand{\dimension}{\mathrm{dim}}
\newcommand{\omni}{\mathcal{E}}
\newcommand{\omnirho}{\rho}
\newcommand{\huai}{\mathrm{I}}
\newcommand{\Hom}{\mathrm{Hom}}

\newcommand{\splinear}{\mathrm{sl}}

\newcommand{\Ker}{\mathrm{Ker}}
\newcommand{\Img}{\mathrm{Im}}

\newcommand{\LDorfman}[1]{   \{      #1  \}       }
\begin{document}

\title{
{Omni-Lie algebroids
\thanks
 {
 Research partially supported by NSFC(19925105) and
CPSF(20060400017).
 }
} }
\author{ Z. Chen and Z.-J. Liu\\
Department of Mathematics and LMAM\\ Peking University,
Beijing 100871, China\\
          {\sf email: chenzhuo@math.pku.edu.cn,\quad liuzj@pku.edu.cn} }

\date{}


\maketitle %\tableofcontents

%\centerline{\bf Dedicated to Professor Murray Gerstenhaber on the
%occasion of his 80th birthday}

\begin{abstract}
 A generalized Courant algebroid structure is  defined on
the direct sum bundle $\dev E\oplus \jet E$, where $\dev E$ and
$\jet E$ are  the gauge Lie algebroid  and the jet bundle of a
vector bundle $E$  respectively.  Such a structure is called an
{\em omni-Lie algebroid} since it is reduced to the omni-Lie
algebra introduced by A.Weinstein if the base manifold is a point.
 We prove that any Lie algebroid structure on $E$ is  characterized by  a Dirac structure
 as the graph of
  a bundle map from
$\jet E $ to $ \dev E$.
\end{abstract}
%\tableofcontents

\section{Introduction}\label{Sec:intro}

The notion of omni-Lie algebras was introduced by Weinstein
\cite{Weinomni} by defining some kind of algebraic structures on $
\gl(V)\oplus V$ for a vector space $V$. Such an algebra is not a
Lie algebra but  all possible Lie algebra structures on $V$ can be
characterized by its Dirac structures. This is why the term {\em
omni} is used here.  The omni-Lie algebra can be regarded as the
linearization of the  Courant algebroid  structure on $TM\oplus
\TStarM$ at a point and are studied from several aspects recently
(\cite{BCG}, \cite{InteCourant}, \cite{UchinoOmni}).

Our purpose is to generalize the  omni-Lie algebra from a vector
space to a vector bundle $E$ in order to characterize all possible
Lie algebroid structures on $E$. It will be seen that the omni-Lie
algebroid is of the form $\omni= \dev{E}\oplus {\jet}{E}$, where
$\dev{E}$ and ${\jet}{E}$ denote the gauge  algebroid and the jet
bundle of $E$ respectively.

An important fact to be discussed in Section \ref{Sec:JetE}  is
that $\dev{E}$ and ${\jet}{E}$ can regarded as  $E$-dual bundles
for each other, i.e., there is a non-degenerate $E$-valued pairing
between these two vector bundles. As a maximal isotropic and
integrable subbundle of $\omni$,  a Dirac structure in the
omni-Lie algebroid turns out to be a Lie algebroid with a
representation on $E$. We prove that  there is a one-to-one
correspondence  between a Dirac structure  coming from
  a bundle map
$\jet E \rightarrow \dev E$  and a Lie algebroid (local Lie
algebra) structure on $E$ when $\mathrm{rank}(E)\geq 2$  ($E$ is a
line bundle).

Let's fix some notations firstly. In this paper, $M$ denotes a
smooth manifold,  $\Id_{\mathcal{C}}$ the identity map for any set
$\mathcal{C}$ and $E\stackrel{q}{\lon}M$ a vector bundle. Before
going to construct an omni-Lie algebroid, let us review some
related notions. Assume that the readers are familiar with Lie
algebroids, which unify the structures of  a Lie algebra and the
tangent bundle of a manifold( please see \cite{Mkz:GTGA} for more
details). The notion of Leibniz algebras was introduced by Loday
\cite{Loday} as follows:
\begin{defi}A Leibniz algebra is a vector space  $L$  with a bilinear operation (not necessarily skew-symmetric) $\LDorfman{\cdot,\cdot}:~  L\times  L\lon  L$
such that the following identity holds.
$$\LDorfman{X,\LDorfman{Y,Z} } =\LDorfman{\LDorfman{X,Y} ,Z}
+\LDorfman{Y,\LDorfman{X,Z} }, \quad  \forall~ X,Y,Z\in  L.
$$
\end{defi}

\noindent\textbf{$\bullet$    Courant algebroids and Dirac
structures.} The Courant bracket on the sections of $T=TM\oplus
\TStarM$ was introduced by Courant \cite{CourantDirac}:
$$\Courant{x_1+\alpha_1,x_2+\alpha_2}= [x_1,x_2]+
\LieDerivation_{x_1}\alpha_2 -\LieDerivation_{x_2}\alpha_1 -\half
\dM(\pairing{x_1,\alpha_2}-\pairing{x_2,\alpha_1}).
$$
For the inner product defined by $
 {(x_1+\alpha_1,x_2+\alpha_2)}_+ =
\half(\pairing{x_1,\alpha_2}+\pairing{x_2,\alpha_1}), $ a Dirac
structure is a  maximal isotropic subbundle $L\subset T$ whose
sections are closed under the Courant bracket. The Dirac
structures include not only Poisson and presymplectic structures,
but also foliations on $M$. A Dirac structure  is also a Lie
algebroid on $M$, whose bracket and anchor are the restrictions of
the Courant bracket and the projection on  $TM$. The properties of
Courant's bracket are the basis for the definition of a Courant
algebroid (\cite{LWXmani},
 \cite{Uchinoremarks}). Recently, several applications of the
Courant algebroid and the Dirac structure have been found in
different fields, e.g., gerbes and generalized complex geometry
(see \cite{BCG}, \cite{GualtieriGeneralizedComplex} for more
details). By introducing a non skew-symmetric  bracket,
\begin{eqnarray}\label{Eqn:Dorfman}
\Dorfman{x_1+\alpha_1,x_2+\alpha_2} &\defbe&
 [x_1,x_2]+\LieDerivation_{x_1}\alpha_2-\LieDerivation_{x_2}\alpha_1+
\dM\pairing{x_2,\alpha_1},
\end{eqnarray}
the pair $(\Gamma(T),\Dorfman{\cdot,\cdot})$ turns out to be  a
Leibniz algebra with the following nice properties.
\begin{eqnarray*}
\rho\Dorfman{e_1,e_2} &=& [\rho(e_1),\rho(e_2)],\\
\Dorfman{e_1,fe_2} &=& f\Dorfman{e_1,e_2}  +\rho(e_1)(f)e_2\,, \\
\Dorfman{e_1,e_1} &=&   \dM {(e_1,e_1)_+},\\
\rho(e_1) {(e_2,e_3)}_+&=& {(\Dorfman{e_1,e_2},e_3)}_+ +
{(e_2,\Dorfman{e_1,e_3})}_+,
\end{eqnarray*}
for all $e_i\in\Gamma(T)$, $f\in \CWM$, where $\rho$: $T\lon TM$
is the projection. Thus a Courant algebroid
%$(T,\Dorfman{\cdot,\cdot}$, $\rho$, $(\cdot,\cdot)_+)$
is  a Leibniz algebroid (\cite{ILMP}) and the twisted bracket
(\ref{Eqn:Dorfman}) is known as the Dorfman
bracket(\cite{Dorfman1987}). This bracket is mentioned in
\cite{LWXmani} and the Leibniz rule is shown in \cite{ROy}.

%\noindent\textbf{$\bullet$  The Omni-Lie Algebras}.
\noindent\textbf{$\bullet$    Omni-Lie algebras.} Motivated by an
integrability problem of the Courant bracket, A. Weinstein gives a
linearization of the Courant bracket at a point \cite{Weinomni}.
Let $V$ be a vector space. Weinstein's bracket is defined on the
direct sum $\huaE=\gl(V)\oplus V$:
$$
\Courant{(\xi_1,v_1),(\xi_2,v_2)}\defbe
([\xi_1,\xi_2],\half(\xi_1(v_2)-\xi_2(v_1))).
$$
This bracket does not satisfy the Jacobi identity. He called
$\huaE=\gl(V)\oplus V$ with the bracket above an \emph{omni-Lie
algebra} because of the following property:
\begin{thm}\label{Thm:Wein}{\rm\cite{Weinomni}}
There is a one-to-one correspondence between a Lie algebra structure
on $V$ and a Dirac structure in $\huaE$ coming from a linear map in
$\Hom(V,\gl(V))$.
\end{thm}
 Here a  Dirac
structure  is a subspace  of   $\huaE$  closed under the bracket
$\Courant{\cdot,\cdot}$ and  maximal isotropic with respect to the
$V$-valued nondegenerate symmetric bilinear form:
$$
 {((\xi_1,v_1),(\xi_2,v_2))}_V\defbe \half(\xi_1(v_2)+\xi_2(v_1)).
$$
That is,  every Lie algebra structure on $V$ can be characterized
by a Dirac structure, which is similar to a Poisson structure on a
manifold.

\noindent\textbf{$\bullet$ Local Lie algebras and Jacobi
manifolds.}

A Lie algebroid is  a special case of local Lie algebras in the
sense of Kirillov \cite{KirillovLocal}. Recall that a local Lie
algebra is a vector bundle $E$ whose   section space $\Gamma(E)$ has
a $\Real$-Lie algebra structure $[\cdot,\cdot]_E$ with the local
property, $\mathrm{supp}[u,v]\subset \mathrm{supp}{u}\cap
  \mathrm{supp}{v}$,  for all $u, v \in \Gamma(E)$, which is also called a Jacobi-line-bundle if $\mathrm{rank}E =
  1$.
In particular, $M$ is called a Jacobi manifold  if the trivial
bundle $M\times\Real $ is a local Lie algebra, which is equivalent
 to that there is a triple $(M,\Lambda, X)$, where $\Lambda$ is a
bi-vector field and $X$ is a vector field on $M$ such that
$[\Lambda,\Lambda]=2 X\wedge \Lambda$ and $[\Lambda,X]=0$
(\cite{LichnerowiczJacobi}).

 In
\cite{WadeConformal}, the  Courant bracket was extended to the
direct sum of the  vector bundle $TM\times \Real$ with its dual
bundle, the  jet bundle $\jet^1(M)\cong \TStarM\times\Real$.  From
this way, it allows one to interpret many structures encountered
in differential geometry in terms of Dirac structures such as
homogeneous Poisson manifolds, Jacobi structures and Nambu
manifolds.

\noindent\textbf{$\bullet$ The jet  bundle of a vector bundle.}
For a vector bundle $E\stackrel{q}{\lon}M$,  one can define its
$1$-jet vector bundle ${\jet^1} E$ by taking an equivalence
relation in $\Gamma(E)$:
$$u_1\sim u_2
 \Longleftrightarrow u_1(m)=u_2(m) ~~\mbox{ and }~~ \dM\pairing{u_{1
},\xi}_m=\dM\pairing{u_{2 },\xi}_m, \quad\forall~
\xi\in\Gamma(\EStar).$$
 $({\jet^1}{E})_m$ is the collection of all equivalence classes.
So any $\mu\in ({\jet^1}{E})_m$ has a representative
$u\in\Gamma(E)$ such that
  $\mu=[u]_m$. There are several equivalent descriptions for
  jet bundles
(see \cite{AlmeidaKumpera} and the references thereof). It is
shown in \cite{CFSecondary} that for any Lie algebroid $E$, each
$k$-order  jet bundle $\jet^k(E)$ inherits a natural Lie algebroid
structure. Let $\p$ be the projection which sends $[u]_m$ to
$u(m)$.  It is known that $\Ker\p \cong  \Hom(TM,E)$ and there is
an exact sequence, referred as the jet sequence of $E$:
\begin{equation}\label{Seq:JetE}
\xymatrix@C=0.5cm{0 \ar[r] & \Hom(TM,E)  \ar[rr]^{~\qquad \e~} &&
                {\jet^1}{E} \ar[rr]^{\p} && E \ar[r]  & 0.
                }
\end{equation}
Moreover, $\Gamma(\jet^1{E})$ is isomorphic to $\Gamma(E) \oplus
\Gamma (T^*M  \otimes E)$ as a $\Real$-vector space
   and any
$u\in \Gamma (E)$ has a lift $[u]\in\Gamma ({\jet^1} E)$ such that
\begin{equation}\label{Eqt:CWMmoduleGammaE}
[fu]=f [u]+ \dM f\otimes u, \quad \forall~ f\in \CWM.
\end{equation}
\noindent\textbf{$\bullet$ The gauge  algebroid of a vector
bundle.}  For a vector bundle $E\stackrel{q}{\lon}M$, its gauge
Lie algebroid $\dev E$ is just the gauge Lie algebroid of the
 frame bundle
 $F(E)$, which is also called the covariant differential operator bundle of $E$ (see \cite[Example
3.3.4]{Mkz:GTGA} and \cite{MackenzieX:1994}).   Here we treat each
element $\frkd$ of $\dev{E}$ at $m\in M$ as a $\Real$-linear
operator $\Gamma(E)\lon E_m$ together with some $x\in T_mM$ (which
is uniquely determined by $\frkd$ and called the anchor of
$\frkd$) such that
$$
\frkd(fu)=f(m)\frkd(u)+x(f)u(m), \, \quad \quad ~~ \forall~
f\in\CWM,
 u\in\Gamma(E).
$$
It is  known that $\dev{E}$ is a  transitive Lie algebroid over
$M$ (\cite{KSK:2002}). The anchor of $\dev{E}$ is given by
$\jd(\frkd)=x$ and the Lie bracket $[\cdot,\cdot ]_{\dev}$ of
$\Gamma(\dev{E})$ is given by the usual commutator of two
operators.
%$$
%[\frkd_1,\frkd_2]_{\dev}=\frkd_1\circ\frkd_2-\frkd_2\circ\frkd_1\,.
%$$
The corresponding  exact sequence,
\begin{equation}\label{Seq:DE}
\xymatrix@C=0.5cm{0 \ar[r] & \gl(E)  \ar[rr]^{\id} &&
                \dev{E}  \ar[rr]^{\jd} && TM \ar[r]  & 0,
                }
\end{equation}
is usually called the  Artiyah sequence.
 The
embedding maps $\e$ and $\id$ in above two exact sequences will be
ignored somewhere if there is no confusion.

\section{The $E$-Duality Between $\dev{E}$ and $\jet^1{E}$}\label{Sec:JetE}

\begin{defi}Let $A$ and $E$ be two vector bundles over $M$. A
vector bundle $B\subset \Hom(A,E)$ is called an $E$-dual bundle of
$A$ if  the $E$-valued pairing $ \conpairing{\cdot,\cdot}_E: ~~
A\times_M B\lon E,$  ~~ $\conpairing{a,b}_E\defbe b(a)$ (where $a\in
A$, $b\in B$) is nondegenerate.
\end{defi}
It is easy to see that $B$ is an $E$-dual bundle of $A$ if and only
if  $A$ is  an $E$-dual bundle  of $B$. In this section we show that
the first jet bundle  $\jet^1{E}$ of a vector bundle
$E\stackrel{q}{\longrightarrow}{M}$ is an $E$-dual of $\dev{E}$ with
some nice properties. Let us now illustrate a procedure that will
yield a new exact sequence from the Artiyah sequence (\ref{Seq:DE}).
 First we consider the dual sequence
$$%\begin{equation}\label{Eqt:DEdual}
\xymatrix@C=0.5cm{
  0 \ar[r] & \TStarM \ar[rr]^{{\jd^*}} && (\dev{E})^* \ar[rr]^{{\id^*}} &&
  E \otimes \EStar \ar[r] & 0. }
$$%\end{equation}
Applying the functor ``$-\otimes E$'', the right end becomes
$E\otimes \EStar\otimes E\cong E\otimes \gl(E)$. Then using the
decomposition $\gl(E)\cong \splinear(E)\oplus \Real \Id_{E}$, we
are able to get a pull-back diagram:
\begin{equation}\label{temp111}
\xymatrix@C=0.5cm{
 0 \ar[r] & \TStarM\otimes E \ar[d]_{\Id}  \ar[rr]^{\epsilon} && \jE
                  \ar[d]_{{\iota}}  \ar[rr]^{\beta } && E \ar[d]_{\huai}  \ar[r]  & 0   \\
 0 \ar[r] & \TStarM\otimes E  \ar[rr]^{{\jd^*}\otimes \Id_{E}} &&
                (\dev{E})^*\otimes E   \ar[rr]^{{\id^*}\otimes \Id_{E}} && E\otimes \splinear(E)~\oplus~ E \ar[r]  & 0.
                }
\end{equation}
Here the right down arrow $\huai$ is the canonical embedding of
$E$ into $E\otimes \splinear(E)~\oplus~ E$ and $\jE$
%(\emph{we will abuse this notion with $\jet^1 E$ later})
 is the pull-back of
(${\id^*}\otimes \Id_{E}$, $\huai$). In other words, for each
$m\in M$,
\begin{eqnarray*}  &&{\jE} =
\set{\nu\in \Hom(\dev{E},E)\,|\,
\nu(\Phi)=\Phi\circ\nu(\Id_{E}),\quad\forall~ ~ ~\Phi\in \gl(E)}.
\end{eqnarray*}
Moreover, the maps $\epsilon$ and $\beta$ in Diagram
(\ref{temp111}) are given respectively by:
$$
\epsilon({\frky})(\frkd)=  \frky\circ\jd(\frkd) ,\quad\forall~ ~
~\frkd\in \dev{E},~ \ {\frky}\in \Hom(TM,E)\,;\quad\quad
$$
$$
\beta(\nu)=\nu (\Id_{E }),\quad\forall~ ~ ~\nu\in  \jE \,.
$$
It is easy to see that ${\jE}$ is   $E$-dual to $\dev{E}$ and it is
called  the \emph{standard $E$-dual bundle  of $\dev{E}$}.
Analogously, one can define the standard $E$-dual bundle  of
${\jE}$, denoted by ${\dE}$, which is given   in the following
pull-back diagram.
\begin{equation}\label{temp6562}
\xymatrix@C=0.5cm{
 0 \ar[r] & \EStar\otimes E \ar[d]_{\Id}  \ar[rr]^{i} && \dE
                  \ar[d]_{{\iota}}  \ar[rr]^{a} && TM \ar[d]_{\huai}  \ar[r]  & 0   \\
 0 \ar[r] & \EStar\otimes E  \ar[rr]^{{\beta^*}\otimes \Id} &&
                (\jE)^*\otimes E   \ar[rr]^{{\epsilon^*}\otimes \Id\ \ ~ ~\qquad ~~~} && TM\otimes \splinear(E)~\oplus~ TM \ar[r]  & 0.
                }
\end{equation}
In other words, $$ {\dE}
\defbe \{\delta\in \Hom( {\jE} ,E )\,|\, \exists~ x\in T M, \, \,~
 \delta(\frky)=\frky(x),~\forall~ ~~ \frky\in \Hom(T
 M, E )\}.
$$
There is  a canonical isomorphism as follows:
\begin{equation}\label{qtyhyutu}
(\cdot)^{\prime}: \dev{E}\cong \dE\ ~~\mbox{ s.t. }~
\frkd^\prime(\nu)=\nu(\frkd),\quad\forall~ ~ ~\frkd\in\dev{E},
~\nu\in\jE.
\end{equation}
Under this isomorphism,  the Artiyah sequence (\ref{Seq:DE}) is
 isomorphic to  the first row in Diagram (\ref{temp6562}).
 Moreover,
there is   a canonical isomorphism  between the jet bundle of $E$
and the standard $E$-dual bundle  of $\dev{E}$.
%For this reason, we introduce an $E$-pairing between $\jE$ and
%$\dev{E}$ taking values in $E$:
%\begin{eqnarray}\label{EpairingjEdevE}
%\conpairing{\nu,\frkd}_E=\conpairing{\frkd,\nu}_E &\defbe&
%{\nu}(\frkd),\quad\forall~ ~ ~\nu\in {\jE},~\frkd\in\dev{E}.
%\end{eqnarray}
%By definition, this pairing possesses the following properties.
%\begin{eqnarray*}
%\conpairing{\nu,\id(\Phi)}_E &=& \Phi\circ \beta(\nu),\quad\forall~ ~~ \Phi\in \gl(E),~\nu\in{\jE};\\
%\conpairing{\epsilon({\frky}),\frkd}_{E} &=& {\frky}\circ
%\jd(\frkd),\quad\forall~ ~ ~\frky\in \Hom(TM,E),~\frkd\in\dev{E}.
%\end{eqnarray*}

\begin{thm}\label{Thm:descriptionofJE}
$\jet^1{E}$ is canonically isomorphic to ${\jE}$.
\end{thm}
\pf We should define a bijective linear map $~\widetilde{\cdot}~:
\jet^1{E}\lon {\jE}$ such that the following diagram commutes.
\begin{equation}\label{Diagram14556}
\xymatrix@C=0.7cm{
 0 \ar[r] & \Hom(TM,E)  \ar[d]_{\Id_{\Hom(TM,E)}} \ar[rr]^{\e} &&
                \jet^1{E} \ar[d]_{\widetilde{\cdot}}  \ar[rr]^{\p} && E \ar[d]_{{\Id_E}} \ar[r]  & 0\\
 0 \ar[r] & \Hom(TM,E)   \ar[rr]^{\epsilon} &&
                \jE   \ar[rr]^{\beta } && E
                  \ar[r]  & 0.
                }
\end{equation}
For each $\mu\in (\jet^1{E})_m$, $\p(\mu)=e\in E_m$, if
$\mu=[u]_m$, for some $u\in \Gamma(E)$, then we define
$\widetilde{\mu}\in ({\jE})_m$ by
\begin{equation}\label{Eqt:tp1}
\widetilde{\mu}(\frkd)=\widetilde{[u]_m}(\frkd)\defbe \frkd u,\quad
\forall~ ~~ \frkd\in \dev{E}_m\,.
\end{equation}
To see that the RHS of (\ref{Eqt:tp1}) is well defined, we need
the following two lemmas.
\begin{lem}\label{Lem:dde}
As Lie algebroids over $M$, $\dev{E}$ and $\dev{\EStar}$  are
isomorphic via ${(\cdot)^{\sim}}$ defined by
\begin{equation}\label{Eqt:dde}
\pairing{\frkd^{\sim}\phi,u}=\jd(\frkd)\pairing{\phi,u}-\pairing{\phi,\frkd
u},\quad \forall~ ~~ \frkd\in \Gamma(\dev{E}),\ u\in\Gamma (E),\
\phi\in\Gamma (\EStar).
\end{equation}
\end{lem}
This fact comes from the isomorphism between the principal frame
bundles $F(E)$ and ${F}(\EStar)$ by sending a frame
  to its dual
frame. For this reason, we can identify $\dev{\EStar}$ with
$\dev{E}$ such that both $\frkd\phi$ and $\frkd u$ make sense,
where  $\frkd$ depends on what is put after it. Notice that, by
this convention, if one treats   $\Phi\in \gl(E)$ as in
$\gl(\EStar)$, it should be  $-\Phi^*$.

\begin{lem}\label{Lem:tp1}
Let $u\in\Gamma(E)$ and suppose that $u(m)=0$. Then for any
$\frkd\in (\dev{E})_m$, one has
\begin{equation}\label{Eqt:frkduuzero}
(\frkd u)^{\uparrow}=u_*(\jd(\frkd))-0_*(\jd(\frkd)).
\end{equation}
Here by $e^\uparrow=\frac{d}{dt}|_{t=0} te\in T_m E$ ($e\in E_m$)
we denote the vertical tangent vector and by $0_*$ we mean the
canonical inclusion of $TM$ into $TE$.
\end{lem}

\pf The RHS of (\ref{Eqt:frkduuzero}) is clearly a vertical
tangent vector of $T_mE$. Thus we need  only to  show that the
results of the two hand sides acting on an arbitrary fiber-wise
linear function, say $l_{\phi}$, for some $\phi\in\Gamma(\EStar)$,
are equal. We see that
\begin{eqnarray*}
(\frkd u)^\uparrow(l_{\phi}) &=&\pairing{\frkd u,
\phi(m)}=\jd(\frkd) \pairing{u,\phi}-\pairing{u(m),\frkd
\phi}\quad (by ~~ (\ref{Eqt:dde} ))\\&=&\jd(\frkd)
\pairing{u,\phi} =
(u_*(\jd(\frkd)))l_{\phi}\\&=&(u_*(\jd(\frkd))-0_*(\jd(\frkd)))l_{\phi}.
\end{eqnarray*}
This completes the proof. \qed

Now we continue to prove Theorem \ref{Thm:descriptionofJE}.
Suppose that $\mu\in (\jet^1{E})_m$ has two representatives $u^1$,
$u^2\in \Gamma(E)$, i.e., $\mu=[u^1]_m=[u^2]_m$. This means that
$$
u^1(m)=u^2(m),\quad u^1_{m*}(x)=u^2_{m*}(x),\quad\forall~ ~~ x\in
T_mM.
$$
 To guarantee   $\widetilde{\mu}$
is well-defined, we need to show that $\frkd(u^1)=\frkd(u^2)$
%\begin{equation}\label{Eqt:tep365464237}
%\frkd(u^1)=\frkd(u^2), \quad
%\end{equation}
 holds for all $\frkd\in (\dev{E})_m$. In fact, let
$v=u^1-u^2\in\Gamma(E)$, which satisfies: $ v(m)=0$ and $
v_{*m}=0_{*m}$. Then the lemma above claims that $(\frkd
v)^{\uparrow}=0$, so that
  $\widetilde{\mu}$ is well-defined. Moreover,
by Definition (\ref{Eqt:tp1}), we  have
\begin{eqnarray*}
\widetilde{\mu} (\Phi)=\Phi (e)=\Phi(\widetilde{\mu}
(\Id_{E_m})),\quad\forall~ ~~ \Phi\in \gl(E_m)\,,
\end{eqnarray*}
and hence $\widetilde{\mu}$ is indeed an element of $({\jE})_m$.
Clearly $~\widetilde{\cdot}~$ is a morphism of vector bundles.

The next step is to prove that (\ref{Diagram14556}) is a commutative
diagram. But we   first need the meaning of the embedding map $\e:
~\Hom(TM,E)\hookrightarrow \jet E$. Take a local trivialization
$E|_{U}\cong U\times E_m$ for some open neighbor $U\cong \Real^k$
($k=\dimension(M)$) containing $m=0$. Then for any $\frky\in
\Hom(TM,E)_m$, define a local section $u\in\Gamma(E|_{U})$ by
\begin{equation}\label{repoffrky}
u(p)=(p,\frky(\overrightarrow{0p})),\quad\forall~ ~ ~p\in U,
\end{equation}
where $\frky(\overrightarrow{0p})$ denotes the tangent vector from
point $0$ to point $p$. Then $u$ is a representative of
$\e(\frky)$.  Following Lemma \ref{Lem:tp1}, we get
\begin{eqnarray*}
\widetilde{\e{(\frky)}}(\frkd)
=\widetilde{[u]}(\frkd)=\frkd(u)=\frky(\jd(\frkd))=\epsilon({\frky})(\frkd),
\quad\forall~ ~~ \frkd\in\dev{E}.
\end{eqnarray*}
This means that  the left square is commutative. The right square
is commutative from the fact that
\begin{eqnarray*}
\beta(\widetilde{\mu}) =
\widetilde{\mu}(\Id_{E_m})=\p(\mu),\quad\forall~ ~ ~\mu\in
\jet^1{E}.
\end{eqnarray*}
Thus the proof of the theorem is finished.  \qed

By means of this theorem  we  identify $\jet^1{E}$ with ${\jE}$
from now on.
%\begin{cor}\label{Cor:jetEquivalentDescription}
Therefore any element $\mu$ of $(\jet^1{E})_m=(\jE)_m$  can be
considered as  a linear map from $(\dev{E})_m$ to $E_m$ satisfying
$$
\mu(\Phi)=\Phi\circ\mu(\Id_{E_m}),\quad\forall~ ~~ \Phi\in \gl(E_m).
$$
Consequently, the jet sequence  (\ref{Seq:JetE}) has a new
interpretation such that the projection $\p$  and the embedding $\e$
of $\Hom(TM,E)$ into $\jet{E}$ are  given by
$$%\begin{equation}\label{Eqt:UpsilonOnfrkd}
\p(\mu)=\mu(\Id_{E_m}), \quad \quad  \e({\frky})(\frkd)\defbe
{\frky}\circ \jd(\frkd),\quad\forall~ ~ ~\frkd\in (\dev{E})_m\,,
$$%\end{equation}
%\end{cor}
%respectively. As mentioned before,  we regard $\TStarM\otimes
%E\cong \Hom(TM,E)\subset \jet{E}$.
By  the canonical isomorphism (\ref{qtyhyutu}),  we can regard
$\dev{E}$  as a subbundle of $\Hom(\jet{E},E)$ and as the standard
$E$-dual bundle  of ${\jE}$. Therefore, there is   an $E$-pairing
between $\jet{E}$ and $\dev{E}$  by setting:
\begin{equation}
%\conpairing{\cdot,\cdot}_E: \quad\jet{E}\times_M\dev{E} (=\dev{E}\times_M\jet{E}) \longrightarrow E,\quad\qquad\\
\label{Eqt:conpairingE} \conpairing{\mu,\frkd}_E~ \defbe
\widetilde{\mu}(\frkd)=\frkd(u),\quad\forall~ ~~ \mu\in
\jet{E},~\frkd\in\dev{E},
\end{equation}
where $u\in \Gamma(E)$ satisfies $\mu=[u]_m$.  Particularly, one
has
\begin{eqnarray}\label{conpairing1}
\conpairing{\mu,\Phi}_E &=& \Phi\circ \p(\mu),\quad\forall~ ~~ \Phi\in \gl(E),~\mu\in\jet{E};\\
\label{conpairing2} \conpairing{{\frky},\frkd}_{E} &=& {\frky}\circ
\jd(\frkd),\quad\forall~ ~~ \frky\in \Hom(TM,E),~\frkd\in\dev{E}.
%\\
%\conpairing{\varsigma,i(\Theta)}_{\EStar}=\Theta\circ
%\p(\varsigma),&\quad&
%\conpairing{e({\frkz}),\frkd}_{\EStar}={\frkz}\circ
%\jd(\frkd);\\
\end{eqnarray}
 Similarly,  $\dev{E^*}$ and $\jet{E^*}$ are $E^*$-dual for each
 other. Meanwhile,  there is also a $\TStarM$-pairing between $\jet{E}$
and $\jet{\EStar}$ given by
$$%\begin{equation}\label{Eqt:conpairingTStarM}
\conpairing{\mu,\varsigma}_{\TStarM}\defbe \dM\pairing{u,\phi},
\quad
 \forall~ \mu\in (\jet{E})_m, ~~ \varsigma\in(\jet{\EStar})_m,
$$%\end{equation}
where $u\in\Gamma(E)$, $\phi\in\Gamma(\EStar)$ satisfy
$\mu=[u]_m$, $\varsigma=[\phi]_m$ respectively.  Combining with
the  isomorphism given in Lemma \ref{Lem:dde}, we can describe the
relations among these four vector bundles  by the following
diagram:
\begin{equation}\label{Fig:4relations}
%(D;A,B,C;M):\quad_{\qDA}
\xymatrix{
 \dev E \ar@{<->}[d]_{\scriptstyle E\mbox{{\small}-dual}} & \ar@{=}[rr]^{\sim}&&
                & \dev \EStar \ar@{<->}[d]^{\scriptstyle\EStar\mbox{{\small}-dual}} &   \\
                %&&&&\\
 \jet E & \ar@{<->}[rr]^{\scriptstyle\TStarM\mbox{{\small}-dual}}
                & & &\jet \EStar    .       }
\end{equation}
The relations above are similar to  the following dual relations,
where  $TE$ and $T\EStar$ are usual dual as two vector bundles
over $TM$.

\begin{equation}\label{relations}
%(D;A,B,C;M):\quad_{\qDA}
\xymatrix{
 T^*E \ar@{<->}[d]_{\scriptstyle \mbox{{\small}}} & \ar@{=}[rr]^{\sim}&&
                & T^*E^* \ar@{<->}[d]^{\scriptstyle\mbox{{\small}}} &   \\
                %&&&&\\
 TE & \ar@{<->}[rr]^{\scriptstyle  \mbox{{\small}}}
                & & & T\EStar    .       }
\end{equation}
 The following diagram is a typical
double vector bundle, by which and the duality theory of double
vector bundles (see \cite{Mkz:GTGA}) one can  explain clearly the
relationship between  Diagrams (\ref{Fig:4relations}) and
(\ref{relations}).
\begin{equation}\label{Fig:TE} \xymatrix{
  \TE \ar[d]_{q_*} \ar[r]^{\pE}
                & E \ar[d]^{q} &   \\
  TM  \ar[r]_{p}
                & M & \ar[l]^{q} E.         }
\end{equation}

\section{Omni-Lie Algebroids and Dirac Structures}\label{Sec:OminLiealgebroid}
Since the gauge Lie algebroid $\dev{E}$ has a natural
representation on $E$, there is the  Lie algebroid cohomology
coming from the complex $(\Gamma(
\Hom(\wedge^\bullet{\dev{E}},E)), ~~\jetd) $. In fact,
 one can
check that $$\jetd u =[u]\in\Gamma (\jet E)
\subset\Gamma(\Hom(\dev E,E)), \quad \forall u \in \Gamma(E).$$
Furthermore, we claim that $\Gamma (\jet E)$ is an invariant
subspace of the Lie derivative $\LieDerivation_{\frkd}$ for any
 $\frkd \in\Gamma(\dev{E})$, which can be defined by the
 Leibnitz rule as follows:
%\begin{eqnarray*}
% \Gamma(\dev{E})\times
%\Gamma(\jet{E})&\lon& \Gamma(\jet{E}),\quad (\frkd,\mu)  \mapsto
%\LieDerivation_{\frkd} \mu,
%\end{eqnarray*}
\begin{eqnarray}\nonumber%\label{frkdmu}
\conpairing{\LieDerivation_{\frkd}\mu,\frkd\pie}_{E}&\defbe&
\frkd\conpairing{\mu,\frkd\pie}_{E}-\conpairing{\mu,[\frkd,\frkd\pie]_{\dev}}_{E},
\quad\forall~ \mu \in \Gamma(\jet{E}), ~
~\frkd\pie\in\Gamma(\dev{E}).
\end{eqnarray}
Actually, it is easy to  check that
\begin{eqnarray}\label{Eqt:pLieDerivation}
\conpairing{\LieDerivation_{\frkd}\mu,\Phi}_E=\Phi\circ\frkd\circ
\p(\mu), ~~~~ \, \forall~ ~~ \Phi\in\Gamma(\gl(E)), ~~~~~~\,
\Longrightarrow  ~~~~ \, \p(\LieDerivation_{\frkd}\mu)=\frkd\circ
\p(\mu).\end{eqnarray} This implies that
$\LieDerivation_{\frkd}\mu \in \Gamma(\jet{E})$ by
(\ref{conpairing1}).

 Now let $ \omni=\dev{E}\oplus
\jet{E}$, which has a nondegenerate symmetric $2$-form from the
$E$-duality:
$$
\ppairingE{\frkd+\mu,\frkr+\nu}\defbe \half(\conpairing{\frkd,\nu}_E
+\conpairing{\frkr,\mu}_E),\quad\forall~ ~~
\frkd,\frkr\in\dev{E},~\mu,\nu\in\jet{E}.
$$
We define the Dorfman bracket on $\Gamma(\omni)$, similar to that
one  mentioned in Section \ref{Sec:intro},
\begin{eqnarray*}
\Dorfman{\frkd+\mu,\frkr+\nu}&\defbe&
[\frkd,\frkr]_{\dev}+\LieDerivation_{\frkd}\nu-\LieDerivation_{\frkr}\mu
+ \jetd\conpairing{\mu,\frkr}_E\,,
\end{eqnarray*}
and call  the quadruple
$(\omni,\Dorfman{\cdot,\cdot},\ppairingE{\cdot,\cdot},\omnirho)$
an {\em omni-Lie algebroid},  where $\omnirho$ is the projection
of $\omni$ onto $\dev{E}$. Comparing with the Courant algebroid,
we can prove that an omni-Lie algebroid has the similar properties
as follows:
\begin{thm}With the  notation above, an omni-Lie algebroid
satisfies the following properties, where $\jd$ is the projection
from $\dev{E}$ to $TM$ in (\ref{Seq:DE}). For all $X,Y,Z\in
\Gamma(\omni)$, $f\in\CWM$,
\begin{itemize}
\item[1)] $(\Gamma(\omni),\Dorfman{\cdot,\cdot})$ is a Leibniz
algebra, \item[2)]
$\omnirho\Dorfman{X,Y}=[\omnirho(X),\omnirho(Y)]_{\dev}$,
 \item[3)]
$\Dorfman{X,fY}=f\Dorfman{X,Y}+(\jd\circ \omnirho(X))(f)Y$,
\item[4)] $\Dorfman{X,X}= \jetd \ppairingE{X,X}$, \item[5)]
$\omnirho(X)\ppairingE{Y,Z}=\ppairingE{\Dorfman{X,Y},Z}+\ppairingE{Y,\Dorfman{X,Z}}$.
\end{itemize}
%(Note that (5) implies (3).)
\end{thm}
  When $E=M\times \Real$,
$\omni\cong(TM\times \Real)\oplus (\TStarM\times\Real)$, the
structure above is studied by Wade in \cite{WadeConformal}.  The
simplest case is that $E=V$, a vector space. Then $\huaE\cong
\gl(V)\oplus V$ and the structure is isomorphic to Weinstein's
omni-Lie algebra. A
 similar  algebraic structure was defined in
\cite{TanLiuGeneralizedLieBialgebras} and  named as a generalized
Lie bialgebra.

\begin{defi}A Dirac structure in the omni-Lie algebroid $\omni$ is a  subbundle
$L\subset \omni$ being maximal isotropic with respect to
$\ppairingE{\cdot,\cdot}$
  and its section space $\Gamma(L)$ is closed under the bracket operation $\Dorfman{\cdot,\cdot}$.
\end{defi}

\begin{pro}
A Dirac structure $L$ is a Lie algebroid with the restricted
bracket and anchor map $\jd\circ \omnirho$. Moreover,
$\omnirho|_L:  L \rightarrow \dev{E}$ gives a representation of
$L$ on $E$.
\end{pro}

This fact is easy to be checked  by the theorem above. Next we are
going to study some special
 Dirac structures and generalize  Theorem \ref{Thm:Wein} of Weinstein from a vector space to a vector bundle. As we shall see, this includes two special cases,
namely the jet algebroid of a Lie algebroid and the $1$-jet
algebroid of a Jacobi manifold. First let us  mention the
following basic fact, for which the proof is merely some
calculations and is ignored.
\begin{lem}\label{Lem:pisharpDiraciff}Given a  bundle map $\pisharp: \jet{E}\lon \dev{E}$,  then  its
graph
$$ L_{\pisharp}=\set{(\pisharp(\mu),\mu)\,|\, \mu\in \jet{E}} \subset
\omni$$ is a Dirac  structure  if and only if
\begin{itemize}\item[1)]
$\pisharp$ is skew-symmetric, i.e., $
\conpairing{\pisharp(\mu),\nu}_E=-\conpairing{\pisharp(\nu),\mu}_E,\quad\forall~
~\mu,\nu\in\jet E; $
 \item[2)]the following equation holds for all
$\mu,\nu\in\Gamma(\jet{E})$.
\begin{equation}\label{Eqt:piEquation}
\pisharp\pibracket{\mu,\nu}=[\pisharp(\mu), \pisharp(\nu)]_{\dev},
\end{equation}
where the bracket $\pibracket{\cdot,\cdot}$ on $\Gamma(\jet{E})$ is
defined by:
\begin{equation}\label{pibracket}
\pibracket{\mu,\nu}\defbe
\LieDerivation_{\pisharp(\mu)}\nu-\LieDerivation_{\pisharp(\nu)}\mu-
\jetd\conpairing{\pisharp(\mu),\nu}_E.
\end{equation}
\end{itemize}
Moreover, such a Dirac structure induces a Lie algebroid
$(\jet{E},\, \pibracket{\cdot,\cdot},\, \jd\circ\pisharp)$.
\end{lem}

\begin{lem}\label{Thm:fourequivalentstatements}For
  the  Lie algebroid $\jet{E}$ induced from a
Dirac structure $L_{\pisharp}$ given above, then the following
statements are equivalent.
\begin{itemize}
\item[1)]$\jd\circ\pisharp\circ\jetd:~ \Gamma(E)\lon \XM$ induces
a  bundle map $E\lon TM$. \item[2)]$\jd\circ\pisharp\circ \e=0$
~{\rm(i.e., $\pisharp (\Img \e)\subset \Img \id$)}.
\item[3)]$\Hom(TM,E)$ is an ideal of $\jet E$. \item[4)]The
quotient Lie algebroid structure  on $E\cong \jet{E}/\Img (\e)$ is
given by
\begin{equation}
\label{Eqt:uvE} \rho_E=\jd\circ\pisharp\circ\jetd, \quad [u,v]_E =
\p \pibracket{\jetd u,\jetd v}=\pisharp(\jetd{u})v,\quad \forall~
~~ u,v\in\Gamma(E).
\end{equation}
\end{itemize}
Here the bundle maps $\e$, $\jd$, $\id$ and $\p$ are given in
exact sequences  (\ref{Seq:JetE}) and (\ref{Seq:DE}).
\end{lem}

\pf 1) $\Rightarrow$ 2) Recall Eqt.(\ref{Eqt:CWMmoduleGammaE}) and
observe that for all $f\in \CWM$, $u\in \Gamma(E)$,
\begin{eqnarray*}
\jd\circ\pisharp\circ\jetd(fu)&=& f\jd\circ\pisharp( \jetd u)+
\jd\circ\pisharp\circ\e (\dM f\otimes u),
\end{eqnarray*}
this implication is obvious.

2) $\Rightarrow$ 3) For any $\frky\in \Gamma(\Hom(TM,E))$ and
$\mu\in \Gamma(\jet E)$, we have
\begin{eqnarray*}
\p\pibracket{\frky,\mu}&=& \p
(\LieDerivation_{\pisharp(\frky)}\mu-\LieDerivation_{\pisharp(\mu)}\frky-\jetd
\conpairing{\pisharp(\frky),\mu}_E)\\
&=&\pisharp(\frky) \p(\mu)-\conpairing{\pisharp(\frky),\mu}_E\quad \quad\mbox{(using (\ref{Eqt:pLieDerivation}))}\\
&=&\conpairing{\pisharp(\frky), \jetd\p (\mu)-\mu}_E\\
&=& (\jetd\p (\mu)-\mu)\circ \jd\circ\pisharp\circ\e (\frky),
\end{eqnarray*}
since $\jetd\p (\mu)-\mu\in \Gamma(\Hom(TM,E))$. So condition 2)
implies that $\p\pibracket{\frky,\mu}=0$, as required.

3) $\Rightarrow$ 4) This implication is   obvious.

4) $\Rightarrow$ 1) For all $u, v\in \Gamma(E)$, $f\in\CWM$, we have
\begin{eqnarray*}
[u,fv]_E&=& \pisharp(\jetd u)(fv)=(\jd\circ\pisharp\circ\jetd u)(f)v+ f\pisharp(\jetd u)v\\
&=&f[u,v]_E+(\jd\circ\pisharp\circ\jetd u)(f)v.
\end{eqnarray*}
This shows that the anchor of the Lie algebroid $E$ should be
$\jd\circ\pisharp\circ\jetd $, which must be a  bundle map. \qed

Suppose that a Lie algebroid $(E,[\cdot,\cdot]_E,\rho_E)$ is
reduced from a bundle map $\pi$ satisfying the conditions in Lemma
\ref{Thm:fourequivalentstatements}, it is not difficult to see
that the anchor $\rho_E: ~E\lon TM$  can be lift to a Lie
algebroid morphism by setting
\begin{equation}\nonumber%\label{Eqt:varpisharp}
\hat{\rho}_E: \jet{E}\lon \dev{(TM)}, \quad
\hat{\rho}_E{[u]_m}=[\rho_E(u),~\cdot~](m),\quad \forall~ ~~
u\in\Gamma(E).
\end{equation}
Moreover, one has the following  commutative diagram such that all
the arrows are Lie algebroid morphisms.

 \begin{eqnarray}\label{Eqt:jetEdiagram}
\xymatrix@C=0.5cm{
 0 \ar[r] & \gl(E)    \ar[rr]^{} &&
                \dev{{E}}    \ar[rr]^{\jd} && TM    \ar[r]  & 0   \\
 0 \ar[r] & \Hom(TM,E)  \ar[u]^{-(\rho_E)^*\otimes{\Id_{E}}}\ar[d]_{-{\Id_{\TStarM}}\otimes {\rho_E}}  \ar[rr]^{} &&
                \jet{{E}}  \ar[u]^{\pisharp}\ar[d]_{\hat{\rho}_E}  \ar[rr]^{\p} && E  \ar[u]^{{\rho_E}}\ar[d]_{{\rho_E}}  \ar[r]  & 0   \\
 0 \ar[r] & \gl(TM)  \ar[rr]^{} &&
                \dev{(TM)}   \ar[rr]^{\jd} && TM \ar[r]  & 0.
                }
 \end{eqnarray}
Now we have two representations of  $ \jet{{E}}$ on $\TStarM\otimes
E\cong \Hom(TM,E)$: (1) the adjoint representation since
$\Hom(TM,E)$ is an ideal of  $ \jet{{E}}$ by Lemma
\ref{Thm:fourequivalentstatements}; (2) the tensor representation of
$\pi$ and $\hat{\rho_E}$ in the above  diagram by identifying
$\dev{(TM)}$ with $ \dev{(T^*M)}$. After some straightforward
computations, we have
\begin{cor}\label{pr}
The above two representations of  $ \jet{{E}}$ on $\TStarM\otimes
E$ are equivalent.
\end{cor}
Conversely, one can get the above  diagram from  a Lie algebroid
$E$ over $M$.

\begin{lem}\label{Lem:LiealgebroidJacobilinetoDirac}
From a Lie algebroid $(E,[\cdot,\cdot]_E,\rho_E)$ one can get
Diagram (\ref{Eqt:jetEdiagram}) by constructing
%a Jacobi-line-bundle $(E,
%[\cdot,\cdot]_E)$, there exits
a Dirac structure $L_{\pisharp}$ in
$\omni$ such that
\begin{equation}\label{Eqt:pisharpjetdu}
\pisharp(\jetd u) = [u,~\cdot~]_E ,\quad\forall~ u\in \Gamma(E).
\end{equation}
Here $[u, \cdot~ ]_{E}$ denotes the corresponding derivation of $E$.
\end{lem}

\pf Since $\Gamma(\jet {E})\cong \Gamma(E) \oplus \Gamma (\TStarM
\otimes E)$, for  a Lie algebroid $(E,[\cdot,\cdot]_E,\rho_E)$,
one can  define a map $\pisharp: \Gamma(\jet{{E}})\lon
\Gamma(\dev{{E}})$ such that (\ref{Eqt:pisharpjetdu}) holds and
for all $f\in \CWM$, $u\in \Gamma(E)$, set
\begin{eqnarray}\nonumber
 \pisharp(\dM f\otimes u) &=& \pisharp(\jetd(fu)-f\jetd
u)\label{Eqn:pisharpdfotimesu} =[fu,~\cdot~]_E -f [u,~\cdot~]_E .
\end{eqnarray}
Then it is easy to check that $\pisharp$ is $\CWM$-linear and hence
it well defines a morphism of vector bundles $\jet E\lon \dev E$.
 By fact that any
section of $\jet{E}$ can be written as a linear combination of the
elements with form $f \jetd{u}$ as well as the property of anchor:
\begin{equation}\label{anchor}
[u,fv]_E=f[u,v]_E+ ((\rho_Eu)f) v, \quad
 ~ ~~ \forall   u,v\in\Gamma(E), ~f\in\CWM,
\end{equation}
 we can check that the
$\pi$-bracket $\pibracket{\cdot,\cdot}$ on $\Gamma(\jet{E})$
defined by (\ref{pibracket}) satisfies the following properties:
\begin{itemize}
\item[1)] $\pibracket{\jetd u_{1},\jetd u_{2}}=\jetd[u_1,u_2]_E$;
\item[2)] $\pibracket{\jetd u_{1},\omega\otimes u_{2}}=
\LieDerivation_{\rho_{E}(u_1)}\omega \otimes u_{2}+ \omega\otimes
[u_1,u_2]_{E}$; \item[3)] $\pibracket{\omega_1\otimes u_1,
\omega_2\otimes u_2}= \pairing{\omega_2,\rho_{E}(u_1)}
(\omega_1\otimes u_2) -\pairing{\omega_1,\rho_{E}(u_2)}
(\omega_2\otimes u_1),$
%\begin{eqnarray*}&&\pibracket{\omega_1\otimes u_1, \omega_2\otimes u_2}\\&
%=& -\pairing{\omega_1,\rho_{E}(u_2)} (\omega_2\otimes u_1)+
%\pairing{\omega_2,\rho_{E}(u_1)} (\omega_1\otimes u_2) ,
%\end{eqnarray*}
\end{itemize}
where $u_i\in\Gamma(E)$, $\omega_i\in \Omega(M)$. It is easy to
see that these relations imply that Eqt. (\ref{Eqt:piEquation}) is
valid and hence $L_{\pisharp}$ is a Dirac structure. Moreover, one
can check that $\jd\circ\pisharp\circ \e=0$. Thus, by Lemma
\ref{Thm:fourequivalentstatements}, the proof is completed.\qed

\begin{rmk}\rm  Actually, it is already  known  to construct the jet Lie algebroid and a representation on
$E$ from  a given Lie algebroid $E$ (see \cite{CFSecondary}). Our
discoveries  include that: (1) the  Lie algebroid structure of
$\jet{E}$ is written clearly in form (\ref{pibracket})  by means of
$\pisharp$ and can be characterized by a Dirac structure; (2) we
find another representation $\hat{\rho_E}$ related to  $\pi$ as
showing in diagram (\ref{Eqt:jetEdiagram}) and Corollary \ref{pr}.
\end{rmk}

The follows are two special cases for ${E}=TM$ with the usual Lie
algebroid structure and $ E= T^*M$ with the Lie algebroid
structure coming from a  Poisson structure.
\begin{cor}There is a canonical  Lie algebroid isomorphism $\hat{\id}_{TM}$: $
\jet{(TM)}\cong \dev{(TM)}$ with  the following commutative
diagram:
$$
\xymatrix@C=0.7cm{
 0 \ar[r] & \gl(TM)^{op}  \ar[d]_{-\Id_{\gl(TM)}}  \ar[rr]^{} &&
                \jet{(TM)} \ar[d]_{\hat{\id}_{TM}}   \ar[rr]^{\p} && TM  \ar[d]_{\Id_{TM}}  \ar[r]  & 0   \\
 0 \ar[r] & \gl(TM)  \ar[rr]^{} &&
                \dev{(TM)}   \ar[rr]^{\jd} && TM \ar[r]  & 0.
                }
$$
\end{cor}

\begin{cor}
Let $(M,\Pi)$ be a Poisson manifold. Then there is a Lie algebroid
morphism $\hat{\Pi}$: $\jet{(\TStarM)}\lon \dev{(TM)}$ such that
the following diagram commutes.
$$
\xymatrix@C=0.7cm{
 0 \ar[r] & \Hom(TM, \TStarM)  \ar[d]_{-\Id_{\TStarM}\otimes \Pi}  \ar[rr]^{} &&
                \jet{(\TStarM)} \ar[d]_{\hat{\Pi}}  \ar[rr]^{\p} && \TStarM  \ar[d]_{\Pi}  \ar[r]  & 0   \\
 0 \ar[r] & \gl(TM)  \ar[rr]^{} &&
                \dev{(TM)}   \ar[rr]^{\jd} && TM \ar[r]  & 0.
                }
$$ In particular, $\jet{(\TStarM)}\cong
\dev{(TM)}$  if $M$ is a symplectic manifold.
\end{cor}
Now we mention the main result of this paper as follows:
\begin{thm}\label{Thm:pisharprankEgeq2}If $\mathrm{rank}(E)\geq 2$, then there is a one-to-one correspondence between
Lie algebroid structures on $E$ and  Dirac  structures in $\omni$
coming from  bundle maps $\jet E\lon \dev E$.
\end{thm}
\pf One direction  is true by Lemma
\ref{Lem:LiealgebroidJacobilinetoDirac}. For the converse
direction, we assume that $L_{\pisharp}$ is a Dirac structure
coming from a skew-symmetric bundle map $\pisharp: \jet E\lon \dev
E$ given in Lemma \ref{Lem:pisharpDiraciff}. We  claim that if
$\mathrm{rank}(E)\geq 2$, then $\jd\circ\pisharp\circ\e=0$. By
equalities (\ref{conpairing1}) and
 (\ref{conpairing2}),  it is seen that $\jd\circ\pisharp\circ\e=0$
is equivalent to
\begin{equation}\label{22}
\conpairing{\pi(\frky),\frkz}_E=0,\quad\forall~ ~~ \frky,\frkz\in
\Hom(TM,E). \end{equation} Taking $\frky =\omega_1\otimes e_1$,
$\frkz = \omega_2\otimes e_2$, for any $\omega_1,\omega_2\in
\TStarM$, $e_1,e_2\in E$, we have
\begin{eqnarray*}
 \conpairing{\pisharp(\omega_1\otimes e_1),\omega_2\otimes
e_2}_E &=&\pairing{\jd\circ\pisharp(\omega_1\otimes
e_1),\omega_2}e_2 =-\pairing{\jd\circ\pisharp(\omega_2\otimes
e_2),\omega_1}e_1.
\end{eqnarray*}
Since $\mathrm{rank}(E)\geq 2$,  $e_1$ and $e_2$ can be
independent so that the coefficients ahead of them must be zero.
This means that (\ref{22}) is true.  Finally, by Lemma
\ref{Thm:fourequivalentstatements}, we know that $E$ has an
induced Lie algebroid structure. \qed
\begin{ex}\rm
Given a Lie algebroid $(E,[\cdot,\cdot],\rho)$,
$\mathrm{rank}(E)\geq 2$,  with the Dirac structure $L_{\pisharp}$
as shown above. For a bundle map $N : E \rightarrow E$, i.e., $N
\in \Gamma(\gl(E)) \subset \Gamma (\dev E)$, the Nijenhuis torsion
of $N$ is defined by, $\forall~ u, v\in
 \Gamma(E)$,
$$
T^N(u,v)\defbe N[u,v]^N-[Nu,Nv],  \quad \mbox{ where }\quad[u,
v]^N=[Nu,v]+[u,Nv]-N[u, v].
$$
We define a twisted bundle map $\pi\circ\hat{N}- ad_N\circ\pi :~
\jet E\lon \dev E,$ where $\hat{N}: ~\jet E\lon \jet E$, $[u]
\mapsto [Nu],$  is the lift of $N$. Then the following three
statements are equivalent.
\begin{itemize}
\item[1)] The graph of  $\pi\circ\hat{N}- ad_N\circ\pi$ is a Dirac
structure. \item[2)]$(E,\,[\cdot,\cdot]^N, ~~\rho\circ N)$ is a Lie
algebroid. \item[3)]
$[T^N(u,v),w]+T^N([u,v],w)+~c.p.~=0,\quad\forall u,v,w\in
\Gamma(E).$
\end{itemize}
In particular, $N$ is a Nijenhuis operator if and only if $T^N =
0$.
\end{ex}\rm
\begin{ex}\rm
Let  $E=M\times V$ be a trivial vector  bundle,  where
$\mathrm{dim}V\geq 2$. In this case,
\begin{equation}\label{directsum}\dev E=(M\times \gl(V))\oplus
TM,\quad \jet E=\Hom(TM,M\times V)\oplus (M\times
V).\end{equation}
 One can check that any skew-symmetric  bundle map $\pisharp: \jet E\lon \dev
 E$ is
determined by a pair of bundle maps $(\theta,\Omega)$, where
$\theta : ~M\times V\lon T M$ and $$\Omega : M\times V\lon M\times
\gl(V), \quad \quad  \Omega (v_1)(v_2)+\Omega
(v_2)(v_1)=0,\quad\forall ~ v_1,v_2\in V,
$$
such  that
$$
\pisharp(\frky,v)=(\Omega (v)- \frky\circ\theta , \, \theta
(v)),\quad\forall ~ (\frky,v)\in  \Hom(TM,M\times V)\oplus
(M\times V).
$$
Moreover, the graph of $\pisharp$ is a Dirac structure if and only
if, ~~ $\forall  v_1,v_2,v_3$ $\in V$ (as constant sections of $E$),
\begin{eqnarray}\label{Eqn:fadf1}
&&\theta\circ\Omega({v_1,v_2})=[ \theta{(v_1)}, \theta{(v_2)}
],\\\label{Eqn:fadf2}
&&\Omega(v_1,\Omega({v_2,v_3}))+L_{\theta(v_1)}\Omega({v_2,v_3})+
c.p. =0.
\end{eqnarray}
The reduced Lie algebroid structure on $E$ is given by   the anchor
$\theta$ and
\begin{equation}\label{trivialbracket}
[u , v ]_E=\Omega({u , v }) +L_{\theta(u)}v-L_{\theta(v)}u\,,
\quad\forall ~ u, v\in C^\infty(M,V).
\end{equation}
In particular,  $\Omega$ is constant if and only if $E$ is an
action Lie algebroid coming from the action of Lie algebra
$(V,\Omega)$ on $M$ by $\theta$. \qed
\end{ex}

From now on we consider the line bundle case. The next example
shows that Equation (\ref{22}) is not always true  for a
skew-symmetric bundle map $\pisharp: \jet E\lon \dev E$ if
$\mathrm{rank}(E) =1$.
\begin{ex}\rm  Suppose that $E$ is the trivial bundle
$M\times \Real$. Then $\jet E\cong \TStarM \times \Real$ and $\dev
E\cong TM\times \Real$. For a skew-symmetric  bivector field
$\Lambda\in \Gamma(\wedge^2 TM)$, define a map $\pisharp: \jet E\lon
\dev E$: $(\xi, t)\mapsto(\Lambda^{\sharp}(\xi), 0)$ by means of the
map $\Lambda^{\sharp}: T^*M \rightarrow TM$. It is easy to see that
 $\pisharp$ is skew-symmetric but
$\jd\circ\pisharp\circ\e=\Lambda^{\sharp}\neq 0$. \qed
\end{ex}

In fact, we can also construct  a Dirac structure for a
Jacobi-line-bundle as doing in Lemma
\ref{Lem:LiealgebroidJacobilinetoDirac}. But it needs more
calculations  because there is no  anchor  in this case.
\begin{lem}\label{Lem:JLB}
From a Jacobi-line-bundle $(E, [\cdot,\cdot]_E)$, one can
construct  a Dirac structure $L_{\pisharp}$ in $\omni$ such that
%\begin{equation}\label{Eqt:pisharpjetdull}
$\pisharp(\jetd u) = [u,~\cdot~]_E ,\quad\forall~ u\in \Gamma(E).$
%\end{equation}
\end{lem}

\pf We still define $\pisharp$ by Eqt.(\ref{Eqt:pisharpjetdu}) and
 show that
$\pisharp$ is really a bundle map and  takes values in
$\Gamma(\dev E)$. Since all calculations are local, without losing
the generality, one can
 assume that $E=M\times \Real$ and identify
$\Gamma(E )$ with $\CWM$. By the result in \cite{KirillovLocal},
 there exists a pair
$(\Lambda,X)$, where $\Lambda$ is a smooth bivector field and $X$
is a smooth vector field   such that
$$
[f,g]_E= \Lambda(\dM　f,\dM　g)+ f X(g)- gX(f),\quad\forall~ ~~
f,g\in \CWM.
$$
Thus we have two expressions of $\pisharp$:
\begin{eqnarray*}
&&\pisharp(\jetd{u})v =
[u,v]_E = \Lambda(\dM u,\dM v)+u X(v)-vX(u);\\
&& \pisharp(\dM f\otimes u)(v) =
 [fu, v]_E-f[u, v]_E
 %\\
%&=& \Lambda( \dM (f u), \dM v) +f u X(v) -v
%X(fu)-f(\Lambda(\dM　u, \dM　
%v)+ u X(v)-v X(u)) \\
= u \Lambda(\dM f,\dM　v)-uv X(f).
\end{eqnarray*}
Therefor  we have, for any $h\in \CWM$,
\begin{eqnarray*}
\pisharp(\jetd{u})(h v) &=& h \pisharp(\jetd{u})v +
(\Lambda^\sharp(\dM u)+uX)(h) v;\\
\pisharp(\dM f\otimes u)(h v) &=& h \pisharp(\dM f\otimes u)v + u
\Lambda^\sharp(\dM f)(h)v,
\end{eqnarray*}
which mean that both $\pisharp(\jetd u)$ and $\pisharp(\dM
f\otimes u)\in\Gamma(\dev{E})$. Using these formulas, it is also
easy to check that $\pisharp$ is a bundle map. Next we  prove that
$L_{\pisharp}$ is a Dirac structure. By some simple calculations,
we get $ \pibracket{\jetd{u},\jetd{v}}= \jetd
[u,v]_E,\quad\forall~ ~~ u,v\in\Gamma(E), $ which implies that
\begin{eqnarray*}
 &&\pisharp\pibracket{\jetd{u},\jetd{v}}(w)
 =  \pisharp(\jetd [u,v]_E)(w)\\&=&[[u,v]_E,w]_E=[[u,w]_E,v]_E +
[u,[v,w]_E]_E\\
&=&
[{\pisharp(\jetd{u}),\pisharp(\jetd{v})}]_{\dev}(w),\quad\forall~
~~ w\in\Gamma(E).
\end{eqnarray*}
Since any local section of $\jet{E}$ can be written as a linear
combination of elements of the form $f \jetd{u}$, the above
equality implies that Eqt.(\ref{Eqt:piEquation}) is valid. \qed

For a line bundle, we have the following theorem analogous to
Theorem \ref{Thm:pisharprankEgeq2}. The difference is that the
quotient structure on $E$ can not be claimed directly since
$\jd\circ\pisharp\circ \e$ maybe not zero in this case.
\begin{thm}\label{Thm:linebundle}For any line bundle $E$, there is a one-to-one
correspondence between local Lie algebra structures on $E$ and
Dirac structures in $\omni$ coming from bundle maps $\jet E\lon
\dev E$. In particular, Dirac structure $L_{\pisharp}$ corresponds
to a Lie algebroid structure of $E$ if and only if
$\jd\circ\pisharp\circ \e=0$.
\end{thm}
\pf   One implication is   shown in Lemma \ref{Lem:JLB}. For the
converse part, let us show that a Dirac structure $L_{\pisharp}$
of the line bundle $E$ determines a local Lie algebra structure
$(E,\,[\cdot,\cdot]_E)$ by setting
\begin{equation}\label{temop562665}
[u,v]_E\defbe \p\pibracket{\jetd u,\jetd v}=
\pisharp(\jetd{u})v,\quad\forall~ ~~ u,v\in\Gamma(E).
\end{equation}
It clearly satisfies the local condition. Moreover, we have
$$\jetd [u,v]_E=\jetd \pisharp(\jetd u)v=
\LieDerivation_{\pisharp(\jetd u)}\jetd v=\pibracket{\jetd u,\jetd
v}.
$$
 To see that $[\cdot,\cdot]_E$ enjoys the Jacobi
identity, we compute, for all $u,v,w\in \Gamma(E)$,
\begin{eqnarray*}
&&[[u,v]_E,w]_E=\pisharp( \jetd[u,v]_E )w=\pisharp( \pibracket{\jetd
u,\jetd v})w\\
&=&[{\pisharp(\jetd{u}),\pisharp(\jetd{v})}]_{\dev}(w)\quad\quad\mbox{(since
$L_{\pisharp}$ is a Dirac structure)}
\\&=&
[u,[v,w]_E]_E-[v,[u,w]_E]_E.
\end{eqnarray*}The last statement of the theorem is already implied
by Lemma\ref{Thm:fourequivalentstatements}.
 \qed

Finally, notice that $\jd\circ\pisharp\circ\jetd : \Gamma(E)
\rightarrow \Gamma(TM)$ is generally not a bundle map for a
Jacobi-line bundle. But it plays a similar role as the anchor of a
Lie algebroid as follows:

\begin{cor}\label{Cor:etqwetr}
For a Jacobi-line bundle and $(E,\,[\cdot,\cdot ]_E )$, one has
$$%\begin{equation}\label{qetrty}
[u,fv]_E=f[u,v]_E+
((\jd\circ\pisharp\circ\jetd{u})f) v, \quad
 ~ ~~ \forall   u,v\in\Gamma(E), ~f\in\CWM .
$$%\end{equation}
\end{cor}
 This equation follows directly from formula (\ref{temop562665}).  In fact, the only obstruction for a Jacobi-line bundle to be a Lie algebroid
is that  $\jd\circ\pisharp\circ \jetd$ is not  a bundle map.


\begin{thebibliography}{999}
\bibitem{AlmeidaKumpera}%
R. Almeida and A. Kumpera, Structure produit dans la cat\'{e}gorie
des alg\`{e}bro\"{\i}des de Lie, \emph{An Acad. Bra. Ci\^{e}nc.}
\textbf{53}(1981), 247-250.

\bibitem{BCG}
H. Bursztyn, G. Cavalcanti, M. Gualteri, Reduction of Courant
algebroids and generalized complex structures,
arXiv:math.DG/0509640.



\bibitem{CourantDirac}
T. Courant, Dirac manifolds,\emph{ Trans. Amer. Math. Soc.}
\textbf{319}(1990), 631-661.
\bibitem{CFSecondary}
M. Crainic and R. L. Fernands, Secondary characteristic classes of
Lie algebroids, \emph{Lect. Notes. Phys.},
\textbf{662}(2005),157-176.
\bibitem{Dorfman1987}
I. Ya. Dorfman. Dirac structures of integrable evolution
equations, \emph{Phys. Lett. A} \textbf{125} (1987), 240-246.

 %\bibitem{Dorfman1993}
 %I. Ya. Dorfman, \emph{Dirac structures and integrability of
 %nonlinear evolution equations,} Wiley, Chichester, 1993.

\bibitem{GualtieriGeneralizedComplex}
M. Gualtieri, \emph{Generalized Complex Geometry,} PhD thesis, St
John's College, University of Oxford, Nov. 2003.
%\bibitem{Hitchin}
%N. Hitchin, \emph{Generalized Calabi-Yau manifolds,}
% arXiv:math.DG/0209099.
\bibitem{ILMP}
R. Ib\'{a}\~{n}es, M. de Le\'{o}n, J.C. Marrero and E. Padr\'{o}n,
Leibniz algebroid asociated with a Nambu-Poisson structure,
\emph{J. Phys. A} \textbf{32}(1999), 8129-8144.
\bibitem{InteCourant}
Kinyon, K.  and A. Weinstein, Leibniz algebras, Courant
algebroids, and multiplications on reductive homogeneous spaces,
\emph{Amer. J. math.,}\textbf{ 123}(2001), 525-550.

\bibitem{KirillovLocal}
A. Kirillov, Local Lie algebras, \emph{{Russian Math. Surveys,}}
\textbf{31}(1976), 55-76.

\bibitem{KSK:2002}
Y. Kosmann-Schwarzbach and K. Mackenzie, Differential operators
and actions of Lie algebroids, {\em Contemp. Math.}, {\bf
315}(2002), 213-233.

\bibitem{LichnerowiczJacobi}
A. Linchnerowicz, Les vari\'{e}t\'{e}s de Jacobi et leurs
alg\'{e}bres de Lie associ\'{e}es, \emph{J. Math. Pures et Appl.},
\textbf{57 }(1978), 453-488.

\bibitem{LWXmani}
Z.-J. Liu, A. Weinstein and P. Xu, Manin triples for Lie
bialgebroids, \emph{J. Diff. Geom.,} \textbf{45}(1997), 547-574.


\bibitem{Loday}
J. L. Loday, Une version non commutative des alg\`{e}bres de Lie:
les alg\`{e}bres de Leibniz, \emph{Enseign. Math. J.}
\textbf{5}(1998), 263-276.

%\bibitem{first}
%K. Mackenzie, \emph{Lie Groupoids and Lie Algebroids in Differential
%Geometry}, LMS Lecture Notes Series, \textbf{124}, Cambridge
%University Press, 1987.
\bibitem{Mkz:GTGA}
K. Mackenzie, \emph{General theories of Lie groupoids and Lie
algebroids}, Cambridge University Press, 2005.


\bibitem{MackenzieX:1994}
 K. Mackenzie and P. Xu, Lie bialgebroids and Poisson groupoids,
\emph{Duke Math. J.}, \textbf{73}(2)(1994),415-452.

\bibitem{ROy}
D. Roytenberg, \emph{Courant algebroids, derived brackets and even
symplectic supermanifolds}, PhD thesis, UC Berkeley, 1999,
arXiv:math.DG/9910078.


\bibitem{TanLiuGeneralizedLieBialgebras}
Y.-J. Tan and Z.-J. Liu, Generalized Lie bialgebras, \emph{Comm.
Alg.} {\bf 26}(7) (1998), 2293-2319.
\bibitem{Uchinoremarks}
K. Uchino, Remarks on the definition of a Courant algebroid,
\emph{Lett. Math. Phys.} \textbf{60}(2002): 171-175.

 \bibitem{UchinoOmni}
 K. Uchino, Courant brackets on noncommutative algebras and omni-Lie
 algebras, arXiv:math.SG/0604101.


\bibitem{WadeConformal}
A. Wade, Conformal Dirac structures, \emph{Lett. Math. Phys.}
\textbf{53}(2000), 331-348.

\bibitem{Weinomni}%
A. Weinstein,  Omni-Lie algebras, Microlocal analysis of the
Schrodinger equation and related topics (Kyoto, 1999). No.
1176(2000), 95-102.



\end{thebibliography}
\end{document}